\documentclass[preprint,12pt]{elsarticle}

\usepackage{amssymb}
\usepackage{algorithm}
\usepackage{algpascal}
\usepackage{algpseudocode}
\usepackage{etoolbox}
\algnotext{EndFor}
\algnotext{EndIf}

 \biboptions{sort&compress}

\algtext*{EndWhile}
\algtext*{EndIf}

\begin{document}

\pagestyle{empty}
\begin{frontmatter}



\title{Two exact algorithms for the generalized assignment problem}


\author{Fatemeh Rajabi-Alni}

\corref{cor1}\ead {fatemehrajabialni@yahoo.com}
\cortext[cor1]{Corresponding author.}

 \address{Department of Computer Engineering, Islamic Azad University,\\North Tehran Branch, Tehran, Iran.\fnref{label3}}

\begin{abstract}

Let $A=\{a_1,a_2,\dots,a_s\}$ and $\{b_1,b_2,\dots,b_t\}$ be two sets of objects with $s+r=n$, \textit {the generalized assignment problem} assigns each element $a_i \in A$ to at least $\alpha_i$ and at most ${\alpha '}_i$ elements in $B$, and each element $b_j \in B$ to at least $\beta_j$ and at most ${\beta '}_j$ elements in $A$ for all $1 \leq i \leq s$ and $1 \leq j \leq t$. In this paper, we present an $O(n^4)$ time and $O(n)$ space algorithm for this problem using the well known Hungarian algorithm. We also present an $O(n^3)$ algorithm for a special case of the generalized assignment, called \textit{the limited-capacity assignment problem}, where $\alpha_i,\beta_j=1$ for all $i,j$.
\end{abstract}

\begin{keyword}

generalized assignment problem\sep limited-capacity assignment problem\sep Hungarian method\sep complete bipartite graph\sep objects with demands and capacities

\end{keyword}

\end{frontmatter}


\section{Introduction}

Given two sets $A$ and $B$, \textit {the assignment problem} aims to optimally assign each object of one set to at least one object of the other set.
The assignment problem has applications in various fields such as computational biology \cite{1}, pattern recognition \cite{2}, computer vision \cite{3}, music information retrieval \cite{4}, and computational music theory \cite{5}. 
Let $A$ and $B$ be two sets with $|A|+|B|=n$, Eiter and Mannila \cite{6} proposed an $O(n^3)$ algorithm for the assignment problem between $A$ and $B$ by reducing it to the minimum-weight perfect matching problem in a bipartite graph. 

In this paper, we consider the generalized assignment between $A$ and $B$, called GA problem, where each point $a_i\in A$ is assigned to at least $\alpha_i$ and at most ${\alpha '}_i$ points in $B$, and each point $b_j\in B$ is assigned to at least $\beta_j$ and at most ${\beta '}_j$ points in $A$, such that sum of the matching costs is minimized. We also present an $O(n^3)$ time algorithm for a special case of the GA problem, where each object must be matched to at least one object.
Schrijver \cite{7} solved the GA problem in strongly polynomial time. We present a new algorithm which computes a generalized assignment between $A$ and $B$ in $O(n^4)$ time using $O(n)$ space. In section \ref{Preliminaries}, we review the basic Hungarian algorithm and some preliminary definitions. In section \ref{newalgorithms}, we present our new algorithms. 

\section {Preliminaries}
\label{Preliminaries}

Given an undirected bipartite graph $G=(A \cup B, E)$, a \textit {matching} in $G$ is a subset of the edges $M \subseteq E$, such that each vertex $v \in A \cup B$ is incident to at most one edge of $M$. Let $Weight(a,b)$ denote the weight of the edge $(a,b)$, the weight of the matching $M$ is the sum of the weights of all edges in $M$, hence $$Weight(M)=\sum_{e \in M}Weight(e).$$
A \textit {max-weight matching} $M$ is a matching that for any other matching $M'$, we have $Weight(M') \le Weight(M)$. 

A path with the edges alternating between $M$ and $E-M$ is called an \textit {alternating path}. Each vertex $v$ that is incident to an edge in $M$ is called a \textit {matched vertex}; otherwise it is a \textit {free vertex}. An alternating path that its both endpoints are free is called an \textit {augmenting path}. Note that if the $M$ edges of an augmenting path is replaced with the $E-M$ ones, its size increases by $1$. 

A \textit {vertex labeling} is a function $l: V \rightarrow \Bbb{R} $ that assigns a label to each vertex $v \in V$. A vertex labeling that in which $l(a)+l(b) \ge Weight(a,b)$ for all $a \in A$ and $b \in B$ is called a \textit {feasible labeling}. The equality graph of a feasible labeling $l$ is a graph $G=(V,E_l)$ such that $E_l=\{(a,b)| l(a)+l(b)=Weight(a,b)\}$. The \textit {neighbors} of a vertex $u \in V$ is defined as $N_l(u)=\{v| (v,u) \in E_l\}$. Consider a set of the vertices $S \subseteq V$, the neighbors of $S$ is $N_l(S)=\bigcup_{u \in S} N_l(u)$.
\newtheorem{lemma}{Lemma}
\begin{lemma}
\label{lem1}
Consider a feasible labeling $l$ of an undirected bipartite graph $G=(A\cup B, E)$ and $S \subseteq A$  with $T=N_l(S)\neq B$, let 
$$\alpha_l=\min_{a_i \in S, b_j \notin T}\{l(a_i)+l(b_j)-Weight(a_i,b_j)\}.$$ If the labels of the vertices of $G$ is updated such that:
$$l'(v)=\left\{ 
\begin{array}{lr}
l(v)-\alpha_l & if \  v \in S 
 \\ 
 l(v)+\alpha_l & if\  v  \in T 
 \\ 
 l(v) & Otherwise 
 \end{array}
\right.,$$ 
then $l'$ is also a feasible labeling such that $E_l \subset E'_l$.
\end{lemma}

\textbf {Proof.} Note that $l$ is a feasible labeling, so we have \newline$l(a)+l(b)\ge Weight(a,b)$ for each edge $(a,b)$ of $E$. \newline After the update four cases arise:
\begin{itemize}
\item $a \in S$ and $b \in T$. In this case $$l'(a)+l'(b)=l(a)-\alpha_l+l(b)+\alpha_l=l(a)+l(b)\ge Weight(a,b).$$ 
\item $a \notin S$ and $b \notin T$. We have $$l'(a)+l'(b)=l(a)+l(b)\ge Weight(a,b).$$
\item $a \notin S$ and $b \in T$. We see that $$l'(a)+l'(b)=l(a)+l(b)+\alpha_l>l(a)+l(b)\ge Weight(a,b).$$
\item $a \in S$ and $b \notin T$. In this situation we have 
$$l'(a)+l'(b)=l(a)-\alpha_l+l(b).$$
Two cases arises: 
\begin{itemize}

\item $l(a)+l(b)-Weight(a,b)=\alpha_l$. So 
$$l'(a)+l'(b)=l(a)-\alpha_l+l(b)=l(a)-l(a)-l(b)+Weight(a,b)+l(b)=Weight(a,b).$$ Hence, $E_l \subset E_{l'}$.  

\item $l(a)+l(b)-Weight(a,b)>\alpha_l$. Obviously 
$$l'(a)+l'(b)=l(a)-\alpha_l+l(b)>Weight(a,b).$$

\end{itemize}

\end{itemize}
\qed

\newtheorem{theorem}{Theorem}
\begin{theorem}
If $l$ is a feasible labeling and $M$ is a Perfect matching in $E_l$, then $M$ is a max-weight matching \cite{8}.
\end{theorem}
\textbf {Proof.} Suppose that $M'$ is a perfect matching in $G$, since each  vertex is incident to exactly one edge of $M'$ we have:
$$Weight(M')=\sum_{(a,b) \in M'} Weight(a,b)\le \sum_{v \in (A \cup B)}l(v).$$ So, $\sum_{v \in (A \cup B)}l(v)$ is an upper bound for each perfect matching. Now assume that $M$ is a perfect matching in $E_l$:
$$Weight(M)=\sum_{e \in M}l(e)=\sum_{v \in (A \cup B)}l(v).$$ It is obvious that $M$ is an optimal matching. 
\qed

In the following, we briefly describe the basic Hungarian algorithm which computes a max-weight perfect matching in an undirected bipartite graph $G=(A \cup B, E)$ with $|A|=|B|=n$. It is obvious that for computing the minimum cost many to many matching using the Hungarian algorithm we must weight each edge $(a_i,b_j)$ by $\frac{1}{Weight(a_i,b_j)}$.

\makeatletter
\expandafter\patchcmd\csname\string\algorithmic\endcsname{\itemsep\z@}{\itemsep=0.5ex plus0.5pt}{}{}
\makeatother

\vspace{0.5cm}

\algsetblock[Name]{Initial}{}{3}{1cm}
\alglanguage{pseudocode}
\begin{algorithm}

\caption{The Basic Hungarian algorithm($A$,$B$)}
\label{BasicHungarian}
\begin{algorithmic}[1]
\Initial \Comment Find an initial feasible labeling $l$ and a matching $M$ in $E_l$
\State Let $l(b_j)=0, \ for \ all \ 1 \le j \le t$
\State $l(a_i)=\max_{j=1}^t Weight(a_i,b_j)\  for\  all \ 1 \le i \le s$
\State $M= \emptyset$
\While {$M$ is not perfect}

\State Select a free vertex $a_i  \in A$ and set $S = \{a_i\}$, $T=\emptyset$
\For{$j\gets 1, n$}
\State $slack[j]=l(a_i)+l(b_j)-Weight(a_i,b_j)$
\EndFor
  \Repeat
     \If {$N_l(S)=T$}
        \State $\alpha_l=\min_{b_j \notin T}slack[j]$
        \State $Update(l)$ \Comment Update the labels according to Lemma \ref{lem1}
\ForAll {$b_j \notin T$}
 \State $slack[j]=slack[j]-\alpha_l$
\EndFor
        \EndIf
         \State Select $u  \in N_l (S)-T$ 
          \If {$u$ is not free}\Comment ($u$ is matched to a vertex $z$, extend the alternating tree)
            \State $S = S \cup \{z\},T = T \cup \{u\}$.
\For{$j\gets 1, n$}
 \State $slack[j]=\min (l(z)+l(b_j)-Weight(z,b_j),slack[j])$
\EndFor
           
           \EndIf
          \Until {$u$ is free}   
    \State $Augment(M)$
\EndWhile
\Return $M$
\end{algorithmic}
 \end{algorithm}

In lines $2$ and $3$, we label all vertices of $B$ with zero and each vertex $a_i \in A$ with $\max_{j=1}^n Weight(a_i,b_j)$ to get an initial feasible labeling. Note that $M$ can be empty. In each iteration of the while loop of lines $5-21$, two free nodes $a_i$ and $b_j$ are matched, so it iterates $O(n)$ times. Using the array $skack[1..n]$, we can run each iteration of this loop in $O(n^2)$ time. The repeat loop runs at most $O(n)$ times until finding a free node $b_j$. In line $11$, we can compute the value of $\alpha_l$ by:
$$\alpha_l=\min_{b_j \notin T}slack[j],$$
in $O(n)$ time. After computing $\alpha_l$ and updating the labels of the vertices, we must also update the values of the slacks. This can be done using:
$$for \ all\  b_j \notin T, slack[j]=slack[j]-\alpha_l.$$ In line $12$, we update the feasible labeling $l$ such that $N_l(S) \neq T$. In line $17$ of Algorithm \ref{BasicHungarian}, when a vertex is moved form $\bar S$ to $S$ the values of $skack[1..n]$ must be updated. This is done in $O(n)$ time. $O(n)$ vertices are moved from $\bar S$ to $S$, so it takes the total time of $O(n^2)$. 

The value of $\alpha_l$ may be computed at most $O(n)$ times in $O(n)$, so running each iteration takes at most $O(n^2)$ time. So, the time complexity of the basic Hungarian algorithm is $O(n^3)$.

\section{The generalized assignment algorithm}
\label{newalgorithms}
In this section, we describe our new algorithm which is based on the well known Hungarian algorithm. Consider two sets $A=\{a_1,a_2,\dots,a_s\}$ and $B=\{b_1,b_2,\dots,b_t\}$ with $s+t=n$. Let $D_A=\{\alpha_1,\alpha_2,\dots,\alpha_s\}$ and $D_B=\{\beta_1,\beta_2,\dots,\beta_t\}$ denote the demand sets of $A$ and $B$, respectively. Let $C_A=\{{\alpha '}_1,{\alpha '}_2,\dots,{\alpha '}_s\}$ and $C_B=\{{\beta '}_1,{\beta '}_2,\dots,{\beta '}_t\}$ be the capacity sets of $A$ and $B$, respectively. Without loss of generality, we assume that $\sum_{i=1}^s{\alpha}'_i\ge\sum_{j=1}^t{\beta}'_j$. 

\begin{figure}
\vspace{-4cm}
\hspace{-11cm}
\resizebox{2.5\textwidth}{!}{%
  \includegraphics{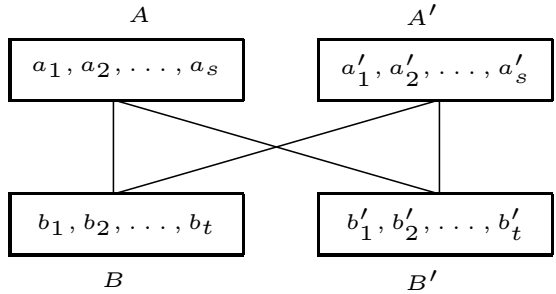}
}
\vspace{-37.5cm}
\caption{Our constructed complete bipartite graph with $\sum_{i=1}^s{\alpha}'_i\ge \sum_{j=1}^t{\beta}'_j$.}
\label{fig:1}       
\end{figure}

The input of our algorithm is the complete bipartite graph that is constructed as follows. Consider the complete bipartite graph $G=(X \cup Y, E)$ where $X=A \cup A'$ and $Y=B \cup B'$ (see Figure \ref{fig:1}). 
A \textit {complete connection} between two sets is a connection that in which each element of one set is connected to all elements of the other set. We show each set of the vertices by a rectangle and the complete connection between them by a line connecting the two corresponding rectangles. 

Given $A= \{a_1, a_2, \dots, a_s\}$ and $B= \{b_1, b_2, \dots, b_t\}$, there exists a complete connection between $A$ and $B$ such that the weight of $(a_i, b_j)$ is equal to the cost of matching the point $a_i$ to $b_j$ for all $1 \le i \le s$ and $1 \le j \le t$. Let $B'= \{b'_1, b'_2, \dots, b'_t\}$ and $A'= \{a'_1, a'_2, \dots, a'_s\}$, each point of $A$ is connected to the all points of $B'$ such that the weight of $(a_i,b'j)$ is equal to the weight of $(a_i,bj)$. There exists also a complete connection between the sets $B$ and $A'$ such that the weight of $(a'_i,bj)$ is equal to the weight of $(a_i,bj)$.

\subsection{The generalized assignment algorithm}

\begin{theorem}
Let $A$ and $B$ be two sets with $|A|+|B|=n$, a generalized assignment between $A$ and $B$ can be computed in $O(n^4)$ time.
\end{theorem}

\textbf {Proof.} We apply our new algorithm, Algorithm \ref{GAA}, on our bipartite graph $G$. Let $Cap(u)$ and $Dem(u)$ denote the capacity and the demand of the vertex $u$; so for all $i,j$ we have $Dem(a_i)=\alpha_i$, $Dem(b_j)=\beta_j$, $Cap(a_i)={\alpha }'_i$, and $Cap(b_j)={\beta }'_j$. 

In our algorithm, a vertex $x$ is free to another vertex $y$ when $x$ is not matched with $y$ in $M$ and has at least one empty capacity.  
So, $a_i \in A$ and $a'_i \in A'$ are called free vertices to a vertex $b$ that are not matched with it in $M$, if $Num(a_i)<Dem(a_i)$ and $Num(a'_i)<Cap(a_i)-Dem(a_i)$, respectively. \newline Also the vertices $b_j$ and $b'_j$ are free to another vertex that is not incident in $M$ to them, when \newline $Num(b_j)<Dem(b_j)$ and $Num(b'_j)<Cap(b_j)-Dem(b_j)$, respectively.

We save the current number of the vertices that are matched to the vertices of $A$, $B$, $A'$, and $B'$ in the arrays $A[1 \dots s]$, $B[1 \dots t]$, $A'[1 \dots s]$, and $B'[1 \dots t]$, respectively; for example $A[i]$ shows the number of the nodes that are matched to $a_i$. The initial values of the arrays is $0$; when a new point is matched to their representing node their values are increased by $1$. Assume that $Num(u)$ returns the number of the vertices that are matched to $u$ so far. So $Num(a_i)=A[i]$, $Num(a'_i)=A'[i]$, $Num(b_j)=B[j]$, and finally $Num(b'_j)=B'[j]$. Note that the procedures $IsFree(u)$ and $IsMatched(u)$ return $True$ if $u$ is free and $u$ is matched, respectively. Therefor, for example in the augmenting path $a,b,c,d$, the vertex $a$ is free to $b$, $b$ is matched to $c$, and $d$ is free to $c$. Now we change the basic Hungarian algorithm as follows.

\algsetblock[Name]{Initialize}{}{4}{1cm}

\alglanguage{pseudocode}
\vspace{1cm}
\begin{algorithm}[H]
\caption{The generalized assignment algorithm}
\label{GAA}
\begin{algorithmic}[1]

\Initialize \Comment Find an initial feasible labeling $l$ and a matching $M$ in $E_l$
\State Let $l(b_j), l(b'_j)=0, \ for \ all \ 1 \le j \le t$
\State $l(a_i)=\max_{j=1}^t (\max(Weight(a_i,b_j),Weight(a_i,b'_j))\  for\  all \ 1 \le i \le s$
\State $l(a'_i)=\max_{j=1}^t Weight(a'_i,b_j)$$\  for\  all \ 1 \le i \le s$
\State Let $M=\emptyset$ 
   \While{$\{w \in B\cup B',\ with\ IsFree(w) \}\neq \emptyset$}

 \State Select $u  \in A\cup A'$ with $IsFree(u)$ 
 \State Set $S = \{u\}, T =\emptyset$
\For{$j\gets 1, t$}
\State $slack[j]=\min(l(u)+l(b_j)-Weight(u,b_j))$

\State $slack[j+t]=\min(l(u)+l(b'_j)-Weight(u,b'_j))$

\EndFor

  \Repeat
      \If {$N_l(S)=T$}
        \State $\alpha_l=\min(\min_{b_j \notin T}slack[j],\min_{b'_j \notin T}slack[j+t])$
        \State $Update(l)$ 
\EndIf
         \State Select $y  \in N_l (S)-T$

         \If {$IsMatched(y)$ }
\Statex\Comment ($y$ is matched to some vertices $z$)
\ForAll{$\{z|(z,y) \in M \ and \ z \notin S \}$}
\State $S \cup \{z\},T = T \cup \{y\}$
\For{$j\gets 1, t$}
 \State $slack[j]=\min (l(z)+l(b_j)-Weight(z,b_j),slack[j])$
\State $slack[j+t]=\min (l(z)+l(b'_j)-Weight(z,b'_j),slack[j+t])$
\EndFor

\EndFor
              \EndIf

          \Until {$IsFree(y)$}   
    \State $Augment(M)$
    \EndWhile

\end{algorithmic}
\end{algorithm}

We first label the vertices of our bipartite graph $G$ using an initial feasible labeling in lines $2-4$. In each iteration of our algorithm, $|M|$ increases by $1$. 
So, our algorithm has $O(n^2)$ iterations with $O(n^2)$ time and runs in $O(n^4)$ time. 

\qed

\subsection{The limited-capacity assignment algorithm} 
Now we present an $O(n^3)$ algorithm for the limited-capacity assignment problem, where each object must be assigned to at least one point of the other set and the capacity of each object is limited.

\begin{theorem}
Let $A$ and $B$ be two sets with $|A|+|B|=n$, a limited-capacity assignment between $A$ and $B$ can be computed in $O(n^3)$ time.
\end{theorem}

\textbf {Proof.} 
We use the bipartite complete graph that is constructed for the generalized assignment problem. We modify the GA algorithm as following.

\algsetblock[Name]{Initialize}{}{4}{1cm}

\alglanguage{pseudocode}
\vspace{1cm}
\begin{algorithm}[H]
\caption{The limited-capacity assignment algorithm (Part I)}
\label{LCA}
\begin{algorithmic}[1]

\Initialize \Comment Find an initial feasible labeling $l$ and a matching $M$ in $E_l$
\State Let $l(b_j), l(b'_j)=0, \ for \ all \ 1 \le j \le t$
\State $l(a_i)=\max_{j=1}^t (\max(Weight(a_i,b_j),Weight(a_i,b'_j))\  for\  all \ 1 \le i \le s$
\State $l(a'_i)=\max_{j=1}^t Weight(a'_i,b_j)$$\  for\  all \ 1 \le i \le s$
\State Let $M=\emptyset$ 
   \While{$\{u \in A,\ with\ IsFree(u) \}\neq \emptyset$}

 \State Select $u  \in A$ with $IsFree(u)$ 
 \State Set $S = \{u\}, T =\emptyset$

\For{$j\gets 1, t$}
\State $slack[j]=\min(l(u)+l(b_j)-Weight(u,b_j))$

\State $slack[j+t]=\min(l(u)+l(b'_j)-Weight(u,b'_j))$

\EndFor

  \Repeat
      \If {$N_l(S)=T$}

      \State $\alpha_l=\min(\min_{b_j \notin T}slack[j],\min_{b'_j \notin T}slack[j+t])$
       
        \State $Update(l)$ 
\EndIf
         \State Select $y  \in N_l (S)-T$

         \If {$IsMatched(y)$ }
\Statex\Comment ($y$ is matched to some vertices $z$)
\ForAll{$\{z|(z,y) \in M \ and \ z \notin S \}$}
\State $S \cup \{z\},T = T \cup \{y\}$
\For{$j\gets 1, t$}
 \State $slack[j]=\min (l(z)+l(b_j)-Weight(z,b_j),slack[j])$

\State $slack[j+t]=\min (l(z)+l(b'_j)-Weight(z,b'_j),slack[j+t])$
\EndFor

\EndFor
              \EndIf

          \Until {$IsFree(y)$}   
    \State $Augment(M)$
    \EndWhile
\algstore{bkbreak}

 \end{algorithmic}
\end{algorithm}

\begin{algorithm}[H]
\caption{The limited-capacity assignment algorithm (Part II)}
\label{LCA}
\begin{algorithmic}[1]
\algrestore{bkbreak}
\While{$\{u \in B,\ with\ IsFree(u) \}\neq \emptyset$}
\State Select $u  \in B$ with $IsFree(u)$ 
 \State Set $S = \{u\}, T =\emptyset$
\For{$i\gets 1, s$}
\State $slack[i]=\min(l(u)+l(a_i)-Weight(a_i,u))$
\State $slack[i+s]=\min(l(u)+l(a'_i)-Weight(a'_i,u))$
\EndFor

  \Repeat
      \If {$N_l(S)=T$}
        \State $\alpha_l=\min(\min_{a_i \notin T}slack[i],\min_{a'_i \notin T}slack[i+s])$

        \State $Update(l)$ 
\EndIf
         \State Select $y  \in N_l (S)-T$

         \If {$IsMatched(y)$ }
\Statex\Comment ($y$ is matched to some vertices $z$)
\ForAll{$\{z|(z,y) \in M \ and \ z \notin S \}$}
\State $S \cup \{z\},T = T \cup \{y\}$
\For{$i\gets 1, s$}
 \State $slack[i]=\min (l(z)+l(a_i)-Weight(a_i,z),slack[i])$
\State $slack[i+s]=\min (l(z)+l(a'_i)-Weight(a'_i,u),slack[i+s])$
\EndFor

\EndFor
              \EndIf

          \Until {$IsFree(y)$}   
    \State $Augment(M)$
\EndWhile
\end{algorithmic}
\end{algorithm}
\qed

\section{Conclusion}
\label{ConclusionSect}
In this paper, we presented an $O(n^4)$ time and $O(n)$ space algorithm for computing a generalized assignment between $A$ and $B$ with total cardinality $n$. In fact, we modified the basic Hungarian algorithm to get a new algorithm, called the generalized assignment algorithm. Then, we construct a bipartite graph $G$ and apply our new algorithm on $G$. We also improved an $O(n^3)$ algorithm for the limited-capacity assignment problem.

\end{document}